\documentclass[sigconf, anonymous=false]{acmart}

\AtBeginDocument{%
  \providecommand\BibTeX{{%
    \normalfont B\kern-0.5em{\scshape i\kern-0.25em b}\kern-0.8em\TeX}}}

\copyrightyear{2021}
\acmYear{2021}
\setcopyright{acmlicensed}
\acmConference[KDD '21]{Proceedings of the 27th ACM SIGKDD Conference on Knowledge Discovery and Data Mining}{August 14--18, 2021}{Virtual Event, Singapore}
\acmBooktitle{Proceedings of the 27th ACM SIGKDD Conference on Knowledge Discovery and Data Mining (KDD '21), August 14--18, 2021, Virtual Event, Singapore}
\acmPrice{15.00}
\acmDOI{10.1145/3447548.3467108}
\acmISBN{978-1-4503-8332-5/21/08}

\usepackage{tabularx,booktabs,multirow,mathtools,rotating}
\usepackage[ruled,vlined]{algorithm2e}
\usepackage{tikz}
\usepackage{multirow}
\usetikzlibrary{arrows,shapes,positioning,shadows,trees,fit,calc,automata,decorations.pathreplacing,arrows.meta}

\usepackage[customcolors]{hf-tikz}

\tikzset{style green/.style={
    set fill color=green!50!lime!60,
    set border color=white,
  },
  style cyan/.style={
    set fill color=cyan!90!blue!60,
    set border color=white,
  },
  style orange/.style={
    set fill color=orange!80!red!60,
    set border color=white,
  },
  hor/.style={
    above left offset={-0.03,0.23},
    below right offset={0.03,-0.12},
    #1
  },
  ver/.style={
    above left offset={-0.14,0.25},
    below right offset={0.12,-0.10},
    #1
  }
}
\begin{document}
\fancyhead{}

\title{Sliding Spectrum Decomposition for Diversified Recommendation}

\author{Yanhua Huang, Weikun Wang, Lei Zhang, Ruiwen Xu}
\affiliation{%
  \institution{Xiaohongshu Inc.}
  \city{Shanghai}
  \country{China}
}
\email{{yanhuahuang, weikunwang, leizhang, ruiwenxu}@xiaohongshu.com}

\begin{abstract}
Content feed, a type of product that recommends a sequence of items for users to browse and engage with, has gained tremendous popularity among social media platforms. In this paper, we propose to study the diversity problem in such a scenario from an item sequence perspective using time series analysis techniques. We derive a method called \textit{sliding spectrum decomposition} (SSD) that captures users' perception of diversity in browsing a long item sequence. We also share our experiences in designing and implementing a suitable item embedding method for accurate similarity measurement under long tail effect. Combined together, they are now fully implemented and deployed in Xiaohongshu App's production recommender system that serves the main Explore Feed product for tens of millions of users every day. We demonstrate the effectiveness and efficiency of the method through theoretical analysis, offline experiments and online A/B tests. 
\end{abstract}

\ccsdesc[500]{Information systems~Recommender systems}
\ccsdesc[500]{Information systems~Information retrieval diversity}
\ccsdesc[500]{Information systems~Similarity measures}

\keywords{Diversified Recommendation; Sliding Spectrum Decomposition; Item Embedding; Determinantal Point Process; CB2CF}

\maketitle

\section{introduction}
\begin{figure}
    \center
    \includegraphics[width=\columnwidth]{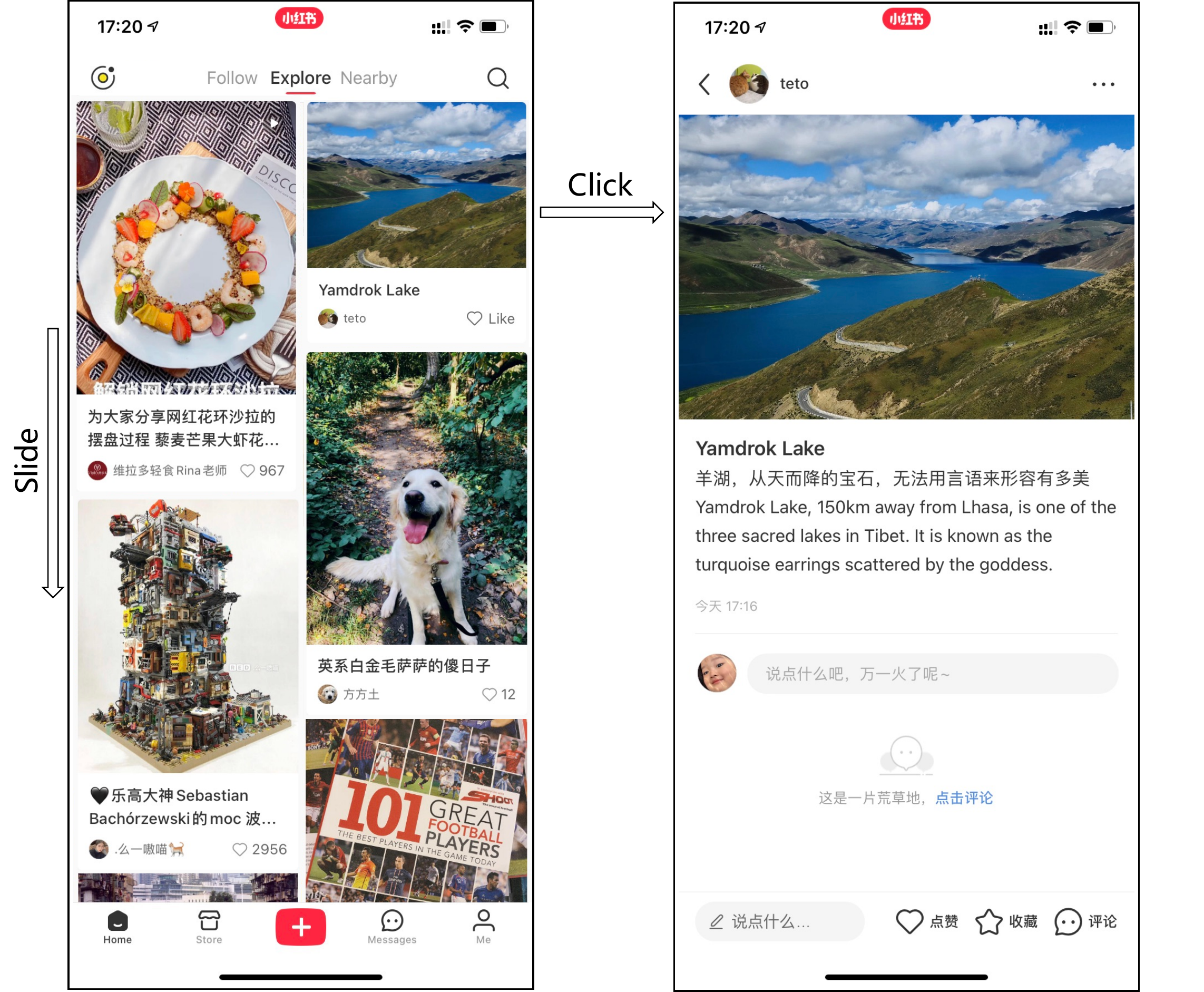}
    \Description[xiaohongshu-screenshot]{A screenshot of Xiaohongshu's Explore Feed product. Left: a 2-columned list of recommended items that users can slide and choose from. Right: the detailed page of an item after user clicks on it.}
    \caption{A screenshot of Xiaohongshu's Explore Feed product. Left: a 2-columned list of recommended items that users can slide and choose from. Right: the detailed page of an item after a user clicks on it.}
    \label{fig:xhs_recsys}
\end{figure}

Content feed, or simply referred to as \textit{feed} in this paper, is a very popular type of social media product that recommends a sequence of items for users to browse and engage with. Instagram's Explore Feed and TikTok's For You Feed are typical examples of such a product. In this paper, we report our work done in Xiaohongshu's Explore Feed product. Xiaohongshu\footnote{https://www.xiaohongshu.com} (meaning \textit{little red book} in Chinese) is a social media platform with more than 100 million monthly active users. Users post their everyday life experiences on fashion, food, travel, reading, fitness, and many more on the platform for fun and inspiration. Each post has both textual and visual contents. In recommeder system's (RecSys) terminology, each post is an item. The platform recommends these items to users via its Explore Feed product as shown in Figure \ref{fig:xhs_recsys}.

\begin{figure}
    \center
    \includegraphics[width=\columnwidth]{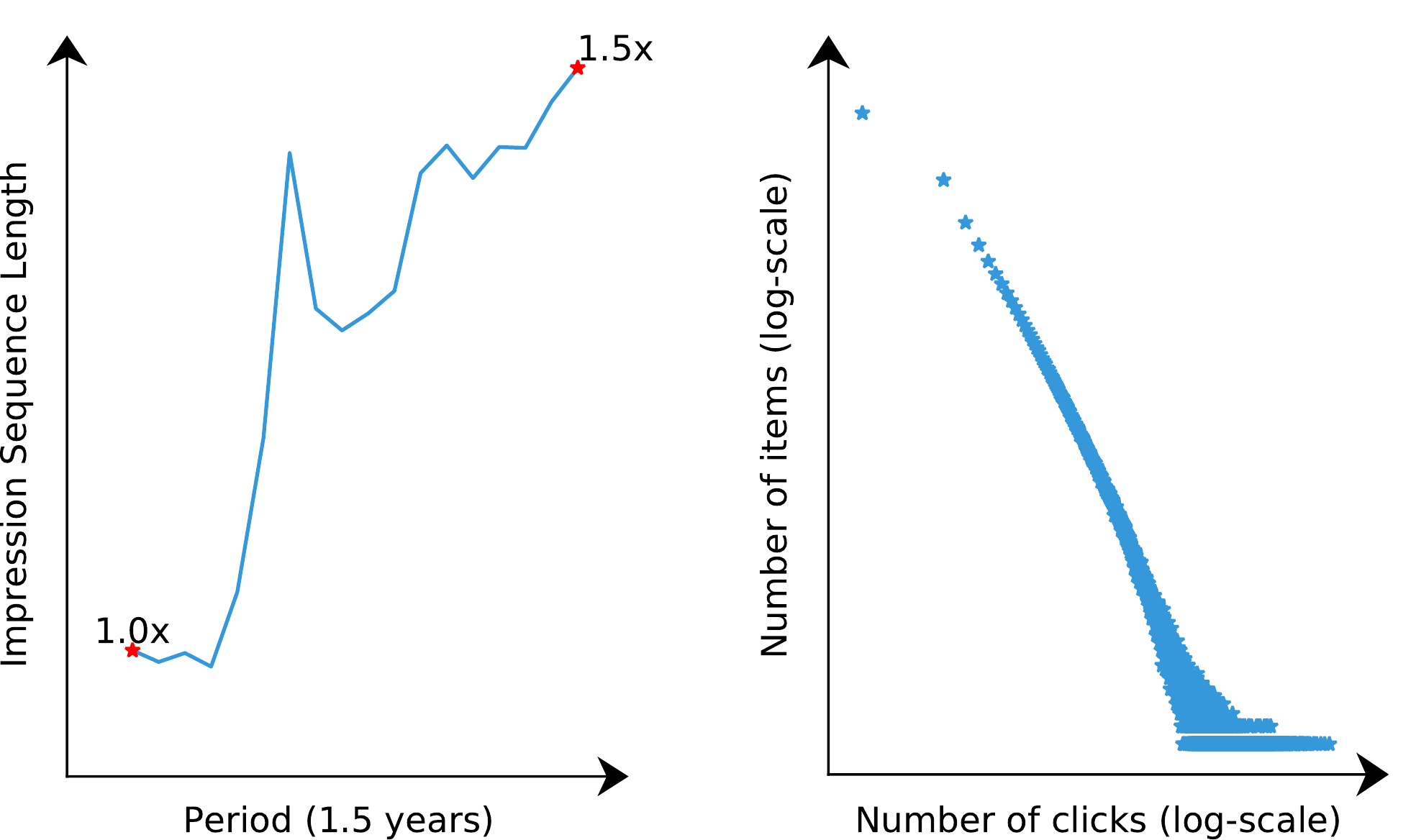}
    \Description[xiaohongshu-data]{Histogram of the number of items over the number of clicks in Xiaohongshu.}
    \caption{Statistics about Xiaohongshu's Explore Feed. Left: The increase of users' average item sequence length in the last 1.5 years. Right: Distribution of the number of items over the number of users' clicks, which clearly shows a long tail effect.}
    \label{fig:xhs_data}
\end{figure}

We identify two common characteristics of these feed products that are strongly related to the problem of diversity in recommendation. First, they allow users to slide down the item list continuously through the phone's current viewing window. Although items in the current viewing window has the most direct impact on the user's perception of diversity, we believe that out-of-window items that the user has already viewed still has a long lasting impact on the user's perception because of user's memory. Through analogy, we found that many existing methods on diversity \cite{carbonell1998use,chen2018fast,wilhelm2018practical} using sliding window based implementations ignore out-of-window items and thus do not fully capture users' perception of diversity. Although enlarging the sliding window may remedy the problem, it nevertheless increases the computation time and in practice hinders its deployment in production systems that has very tight latency requirement. This problem is exacerbated in the feed scenario because users nowadays tend to view many items on feed applications. Figure \ref{fig:xhs_data} (Left) shows that the average length of user viewed item sequence of Xiaohongshu Explore Feed increased about 50\% in the last 1.5 years. Existing algorithms on diversified recommendation with a sliding window implementation will ignore many items in this long item sequence scenario. To solve this problem, we propose to study the diversified recommendation problem from an item sequence perspective using time series analysis techniques. We derive a method called \textit{sliding spectrum decomposition} (SSD) in this direction. It stacks multiple sliding windows into a trajectory tensor. Decomposing the tensor leads to a generalized definition of diversity that takes the out-of-window items and thus the entire item sequence into consideration. A greedy inference procedure is designed to compute the diversity term and is shown to be more efficient than the state-of-the-art methods \cite{chen2018fast}.

Second, as in many recommender systems, the long tail effect emerges in these feed products. A small number of items gets most of the user engagements while a large number of items has no or very few user interactions. Figure \ref{fig:xhs_data} (Right) shows the effect in Xiaohongshu's Explore Feed. One of the key components in diversity algorithms is the measurement of item similarity. The long tail effect makes collaborative filtering (CF) \cite{linden2003amazon,koren2009matrix,rendle2012bpr} based similarity measurement less effective due to the sparseness of user engagements on many items. One alternative is to rely on content based  (CB) methods. However, purely content based similarity measurement also has some drawbacks. Collecting a large number of training data for content based similarity model is very time consuming and the training data tends to bias towards the annotators' view of item similarity instead of the end users' \cite{cremonesi2013user}. Inspired by the CB2CF work \cite{barkan2019cb2cf}, we design a similar strategy that forces a model to generalize from item content to learn the users' view of item similarity as implicitly expressed in their item engagements. We also name this strategy as CB2CF although the actual method is different from \cite{barkan2019cb2cf}. Following this strategy, we design and train a siamese network \cite{bromley1994signature} to learn the item embedding. The embeddings are then adapted and used for similarity computation in the SSD method.

We summarize our contributions as follows:
\begin{itemize}
    \item Unlike many previous methods \cite{carbonell1998use,chen2018fast,wilhelm2018practical}, we study the recommendation diversity problem from an item sequence perspective using time series analysis techniques. We believe this perspective aligns more with the feed scenario where a sequence of items is recommended to users.
    \item Following this direction, we derive a method called \textit{sliding spectrum decomposition} (SSD). It incorporates out-of-window items and thus the entire item sequence into consideration for diversity. In addition, it is computationally more efficient compared to the state-of-the-art method \cite{chen2018fast}.
    \item We share our experiences on using CB2CF strategy to obtain an item embedding that can accurately measure item similarity under long tail effect. We describe how to adapt the embedding vector to work with SSD and how the embedding vector performs in our production data. 
    \item We demonstrate the effectiveness and efficiency of the proposed methods through theoretical analysis, offline experiments and online A/B tests.
\end{itemize}


\section{Related Work}

\subsection{Diversified Recommendation}
There are mainly two perspectives on the diversity in recommender systems: aggregation and individual. Aggregation considers the diversity across all users, where the goal is usually to promote the coverage of the recommender\cite{herlocker2004evaluating, ge2010beyond}. On the other hand, individual, also the one we study in this work, focuses on the diversity for a given user, where we usually want to achieve the best trade-off between quality and diversity\cite{carbonell1998use,abdool2020managing,wilhelm2018practical,ashkan2015optimal}.

Under the individual perspective, diversified recommendation is usually formulated by an optimization problem with an objective considering both quality and diversity. Several works discussed how to construct and solve this optimization problem. The work in \cite{carbonell1998use} provided a marginal-based greedy algorithm, also known as maximal marginal relevance (MMR), where the score of an item is a retrieval quality term minus a penalty term representing its similarity to previously selected items. \cite{qin2013promoting} proposed to maximize the relevance plus an entropy-like diversity term. Recently, many works adapt the determinantal point process (DPP) to the diversified recommendation. DPP is a probabilistic framework, which was first introduced to describe the distributions in thermal equilibrium, characterized as the determinants of some function \cite{macchi1975coincidence}. Applications on recommendation usually use the L-ensemble DPP, where the function is a symmetric matrix, also known as the kernel matrix, representing qualities and pairwise similarities. In this paper, DPP is always referred to as L-ensemble DPP. In \cite{chen2018fast}, the authors provided a fast greedy algorithm to address the computational complexity issue of DPP. The work in \cite{wilhelm2018practical} proposed a practical method that learns the kernel matrix from past interactions together with an approximate greedy inference. To improve efficiency in practice, the sliding window is a key component in these methods \cite{kapt2010result,wilhelm2018practical}. During the greedy inference with a sliding window, all of these works entirely ignores out-of-window items. However, it is more in line with users' perception to maintain the information of these items properly. This work models the recommendation sequence as a time series observed by the user. By taking advantage of the time series analysis technique, our method is able to consider multiple windows to model the diversity of the entire sequence.

\subsection{Item Similarity}
In recommender system, item-to-item recommendation is a key component where most similar items of users' liked items are presented. Techniques that aim to compute item similarities for this purpose are often characterized as either Collaborative Filtering (CF) or Content Based (CB). CF algorithms, which builds a similar items table from users' implicit feedback, are commonly used for a variety of personalization tasks~\cite{linden2003amazon,su2009survey}. 
Content Based (CB) approaches, on the other hand, ignore user's historical behavior and computes item similarities based on content information, such as item description, image, etc. 

In the existing works of diversified recommendations, both CB and CF techniques have been explored to measure item similarity. For instance, the work in~\cite{wilhelm2018practical} suggested to use Jaccard distance between item tokens as the distance function. In~\cite{chen2018fast}, the authors created item representation with matrix factorization technique from user feedbacks. In general, CF approaches provide more accurate similarity measurement compared with CB methods~\cite{slaney2011web}. However, CF algorithms are limited when there is no or few feedbacks on items, i.e., long tail or new items, which constitutes the major part of items in social media platforms. Therefore, many attempts have been made to solve this issue with content based features. For instance, in~\cite{barkan2019cb2cf} the authors proposed a CB2CF method to learn a mapping from content features to collaborative filtering embeddings to address cold-start item recommendation. Further in~\cite{barkan2020neural}, the authors improved the model by utilizing a multiview neural attention mechanism. However, the attention model is not suitable to be used to measure item similarity for diversified recommendation tasks since it is pair-wise model and computationally prohibitive given a large number of paired items to be evaluated. In this paper, inspired by the work in~\cite{barkan2019cb2cf}, we propose a suitable CB2CF item embedding to work with SSD. We differ from the approach in that we do not map item contents into CF vectors. Instead, we force the model to learn from item content to infer a strengthened similarity signal produced by ItemCF method \cite{linden2003amazon}.

\section{Problem Setup}
\begin{figure}
    \begin{center}
        \begin{tikzpicture}[node distance=2cm,>=stealth, stuff/.style={rectangle, draw, inner sep=0pt, thick, minimum size=4.5em, align=center}]
        \draw 
            node [stuff, ] (retrieval) {Retrieval}
            node [stuff, right of=retrieval] (ranking) {Ranking}
            node [stuff, right of=ranking] (policy) {Policy}
            node [stuff, right of=policy] (user) {User}
        ;
        \draw[-{Latex[width=2mm]}, thick] (retrieval.east) -- (ranking.west);
        \draw[-{Latex[width=2mm]}, thick] (ranking.east) -- (policy.west);
        \draw[-{Latex[width=2mm]}, thick] (policy.east) -- (user.west);        
        \end{tikzpicture}
    \end{center}
    \Description[structure-of-industrial-recsys]{Structure of Industrial Recommender System}
    \caption{A typical structure of industrial recommender systems.}
    \label{fig:structure_recsys}
\end{figure}
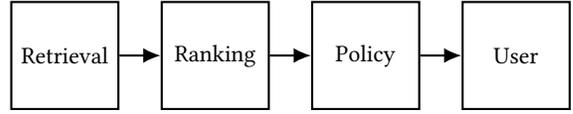

Figure \ref{fig:structure_recsys} demonstrates a typical structure of industrial recommender systems \cite{wilhelm2018practical,wang2020cold}. The system first retrieves candidate items from item databases, followed by a ranking module to measure a pointwise quality, i.e., a prediction of how much the utility that a particular item will provide to the user, for each item. Finally, the high-quality items will be sent to the policy module for further selection and reranking to build the final recommendation.

In this paper, we consider the diversified feed recommendation problem, where the policy module needs to select an item sequence $\{i_1, ..., i_T\}$ from $N$ candidate items $Z = \{1, 2, ..., N\}$ generated by the ranking module. The selection procedure should consider the diversity among items as well as their qualities. In our settings, for each item $i$, the ranking module provides an estimation $r_i$ for its quality, and the policy module has access to its $d$-dimensional embedding $\boldsymbol{v}_i \in \mathbb{R}^d$. We require that the inner product of two items' embeddings represents their similarity. Moreover, nowadays, the retrieval and ranking module usually use complex deep neural networks, posing a strict latency requirement to our solution in the policy module.

\section{SSD}
In this section, we first formulate our proposed method, \textit{sliding spectrum decomposition} (SSD), for diversified feed recommendation. Then we introduce an efficient greedy inference method for SSD to meet the low latency requirement of the policy module in production system. Finally, we provide an in-depth analysis of SSD and show its deep connection with the DPP method \cite{chen2018fast}.

\subsection{Formulation}  
In feed recommendation, a user browses a long sequence of items and interacts with them. Our goal is to construct a high-quality and diverse item sequence in the policy module of RecSys in this case. We solve the problem by answering two questions: 1) how to measure the diversity in such a long sequence; 2) how to construct an objective considering the overall quality and diversity.

\begin{figure}
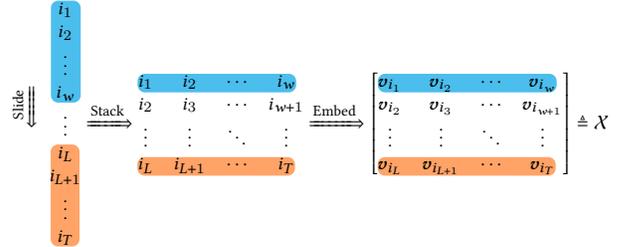

  \resizebox{.8\columnwidth}{!}{
  \begin{minipage}{\columnwidth}
  \begin{displaymath}
   \hspace{-10mm}
   \rotatebox{90}{$\xLeftarrow{\text{Slide}}$}
    \begin{matrix}
      \tikzmarkin[ver=style cyan]{lw 1} i_1\\
      i_2 \\
      \vdots \\
      i_w \tikzmarkend{lw 1} \\
      \vdots \\
      \tikzmarkin[ver=style orange]{lw 2} i_{L} \\
      i_{L+1} \\
      \vdots \\
      i_T \tikzmarkend{lw 2}\\
    \end{matrix} \xRightarrow{\text{Stack}} 
    \begin{matrix}
      \tikzmarkin[hor=style cyan]{mw 1} i_1 & i_2 & \cdots & i_w \tikzmarkend{mw 1} \\
      i_2 & i_3 & \cdots & i_{w+1} \\
      \vdots & \vdots & \ddots & \vdots \\
      \tikzmarkin[hor=style orange]{mw 2} i_{L} & i_{L+1} & \cdots & i_T \tikzmarkend{mw 2}\\
    \end{matrix} \xRightarrow{\text{Embed}} 
    \begin{bmatrix}
      \tikzmarkin[hor=style cyan]{rw 1} \boldsymbol{v}_{i_1} & \boldsymbol{v}_{i_2}  & \cdots  & \boldsymbol{v}_{i_w} \tikzmarkend{rw 1}  \\
      \boldsymbol{v}_{i_2} & \boldsymbol{v}_{i_3}  & \cdots  & \boldsymbol{v}_{i_{w+1}} \\
      \vdots & \vdots  & \ddots   & \vdots  \\
      \tikzmarkin[hor=style orange]{rw 2} \boldsymbol{v}_{i_L} & \boldsymbol{v}_{i_{L+1}}  & \cdots\  & \boldsymbol{v}_{i_T} \tikzmarkend{rw 2} \\
    \end{bmatrix}\triangleq\mathcal{X}
  \end{displaymath}
  \end{minipage}
  }
  \caption{From user's observation to the trajectory tensor.}
  \Description{Constructing the trajectory tensor from user's observation.}
  \label{fig:construct_tensor}
\end{figure}

For the first question, to align with users' perception, let us analyze the item sequence from a user's standpoint. Since the feed recommender usually provides a long sequence of items, a user's current viewing window can only take up a part of the whole sequence. As the user browses items continuously, it is natural to regard the feed as a time series from a user's observation. In time series analysis, it is common to involve a sliding window to represent temporal resolution. Similarly, in recommender system the sliding window for the user appears as the current screen of a mobile phone or a cache area in the user's mind. Assume that this window has a fixed size, marked as $w$, we can process a user's observation as in Figure ~\ref{fig:construct_tensor} by three steps. First, a window slides throughout the raw item sequence. Then multiple windows of items are stacked together in the form of an item matrix. Finally, we map each item to its $d$-dimensional item embedding, obtaining the following \textit{trajectory tensor} $\mathcal{X} \in \mathbb{R}^{L \times w \times d}$:
\begin{equation}
\mathcal{X} = \begin{bmatrix}
\boldsymbol{v}_{i_1} & \boldsymbol{v}_{i_2}  & \cdots   & \boldsymbol{v}_{i_w}   \\
\boldsymbol{v}_{i_2} & \boldsymbol{v}_{i_3}  & \cdots   & \boldsymbol{v}_{i_{w+1}}  \\
\vdots & \vdots  & \ddots   & \vdots  \\
\boldsymbol{v}_{i_{L}} & \boldsymbol{v}_{i_{L+1}}  & \cdots\  & \boldsymbol{v}_{i_T}  \\
\end{bmatrix},\end{equation}
where $L=\max(1, T - w + 1)$. When $d = 1$, $\mathcal{X}$ is exactly the \textit{trajectory matrix} in singular spectrum analysis (SSA) for univariate time series \cite{broomhead1986extracting}. In time series analysis, SSA is a powerful technique that has been widely applied to various fields, e.g., multivariate statistics, nonlinear dynamical systems, and signal processing\cite{vautard1989singular}. In traditional time series analysis, a complex time series usually consists of several regular components. For example, with the continuous progress of agriculture, the grain production shows an upward trend but is also influenced by seasons, i.e., the time series of the grain production is the sum of the trend and the seasonality. SSA is such a technique that can decompose a time series into various orthogonal components, where these components' weights are represented by the singular values through the singular value decomposition of the trajectory matrix.

In recommendation scenario, we generalize the trajectory matrix to third-order case by regarding $d$-dimensional item embeddings as multivariate observations. Following SSA, we perform the singular value decomposition of $\mathcal{X}$ \cite{kolda2001orthogonal,tucker1966some}:
\begin{equation}
\mathcal{X} = \sum_{\sigma_{ijk} > 0} \sigma_{ijk} \boldsymbol{u}_i^{(1)} \otimes \boldsymbol{u}_j^{(2)} \otimes \boldsymbol{u}_k^{(3)},
\end{equation}
where $\boldsymbol{u}_i^{(1)} \in \mathbb{R}^{L}$, $\boldsymbol{u}_j^{(2)} \in \mathbb{R}^w$, $\boldsymbol{u}_k^{(3)} \in \mathbb{R}^d$ are columns of the orthogonal decomposition matrices of $\mathcal{X}$, and $\otimes$ refers to the outer product. In RecSys, the orthogonal components in the user's observation can be seen as the orthogonal directions presented by the contents of items, and the singular values thus refer to these directions' weights in the user's perception of diversify.

After the above processing, the remaining problem is how to define the diversity from this decomposition. Let us first consider a simple case where there is no overlap between any pair of windows, i.e., windows are independent of each other when calculating the diversity. We thus can only focus on a single row of $\mathcal{X}$, degenerating the trajectory tensor to a matrix. Assume that items are embedded in an inner product space, i.e., the inner product of a pair of items can represent the similarity. It is natural to define the diversity by the volume of the hyper-parallelepiped spanned by these items, where diverse items span a larger volume because their embeddings are more orthogonal. Note that one of the calculation methods for the volume of a matrix is to use the cumulative product of singular values\cite{trefethen1997numerical}. We thus generalize this method to define the volume of the third-order tensor $\mathcal{X}$ as follows:
\begin{equation}
\label{eq:diversity_def}
\prod_{\sigma_{ijk} > 0}\sigma_{ijk}.
\end{equation}
Note that in the third-order tensor $\mathcal{X}$, SSD combines multiple windows together from users' perception when browsing the entire sequence. $\mathcal{X}$'s volume thus represents the diversity based on the whole sequence as well as the sliding window. As a consequence, we define the diversity of the entire sequence by Eq.~\eqref{eq:diversity_def}.

A remaining question is how to organize the diversity and the quality in the optimization, we propose to sum them directly:
\begin{equation}
\label{eq:ssd_obj}
\max_{\{i_1, ..., i_T\} \subset Z} \sum_{t=1}^T r_{i_t} + \gamma \prod_{\sigma_{ijk} > 0}\sigma_{ijk},
\end{equation}
where $\gamma$ is a hyper-parameter to adjust the trade-off between quality and diversity. We show the rationality and insight of Eq.~\eqref{eq:ssd_obj} from the standpoint of a recommender. On one hand, an industrial recommender usually computes the quality scores to represent the expected utility of the user, such as video view time or engagements. On the other hand, diversity is sometimes regarded as an exploration to discover more about the user \cite{santos2010exploiting,teo2016adaptive}. Therefore the trade-off between quality and diversity is an exploit-explore for the recommender. Especially, consider the item sequence as a bandit with the following Gaussian parameterized reward:
\begin{equation}
\mathcal{N}(\sum_{i=1}^T r_i, \prod_{\sigma_{ijk} > 0}\sigma_{ijk}^2),
\end{equation}
where a better diversity provides a higher variance in the item sequence\cite{khrouf2013hybrid,eskandanian2017clustering}. It is also in consistency with the volume definition of diversity as a larger volume intuitively represents a larger variance. Note that the upper confidence bound of a Gaussian variable is linearly proportional to its standard deviation. With the principle of optimism in the face of uncertainty, Eq.~\eqref{eq:ssd_obj} is a UCB-type arm-selection strategy \cite{auer2002finite,li2010contextual}. 

SSD regards the whole item sequence as a time series observed by the user, and decomposes its sliding representation with spectrum analysis. As a result, we call our proposed method as \textit{sliding spectrum decomposition}.


\subsection{Greedy Inference}
\label{sec:greedy_inference}
We have formulated diversified feed recommendation problem as a maximum optimization problem in Eq. \eqref{eq:ssd_obj}. Optimizing the diversity term is a combinatorial tensor decomposition problem, which is proved to be NP-hard \cite{hillar2013most}. In this section, we provide a fast greedy inference algorithm to address the efficient inference issue in practice. 


In the $t$-th step of the greedy inference, the previous $t - 1$ items have been already selected and we need to select an item from the candidate set $Z \setminus I_{1:t}$ with respect to Eq. \eqref{eq:ssd_obj}, where $I_{p:q}$ refers to $\{i_p, i_{p+1}, ..., i_{q - 1}\}$.

Let us begin with time steps $t \le w$. When $t = 1$, it is straightforward to select the item with the highest quality. In other cases, it is unnecessary to know the actual value of each singular value, i.e., we only care about the cumulative product. Recall that the cumulative product is equal to the volume in geometry. When $t = 2$ and we consider the candidate item $j$, the volume is the area of the parallelogram spanned by $i_1$ and $j$: $\Vert\boldsymbol{v}_{i_1}\Vert \Vert\boldsymbol{v}_j \sin(\boldsymbol{v}_{i_1}, \boldsymbol{v}_j) \Vert$, where $\Vert\cdot\Vert$ refers to the L2-norm. Note that $\boldsymbol{v}_j \sin(\boldsymbol{v}_{i_1}, \boldsymbol{v}_j)$ is actually the orthogonalized $\boldsymbol{v}_j$, marked as $\boldsymbol{v}_{j}^\perp$, with respect to $\boldsymbol{v}_{i_1}$:
\begin{equation}
\boldsymbol{v}_{j}^\perp = \boldsymbol{v}_j - \frac{\langle \boldsymbol{v}_{j}, \boldsymbol{v}_{i_1} \rangle}{\langle \boldsymbol{v}_{i_1}, \boldsymbol{v}_{i_1} \rangle}\boldsymbol{v}_{i_1},
\end{equation}
where $j$'s projection on $i_1$ is removed from $\boldsymbol{v}_{j}$. When $t = 3$, the volume of the corresponding parallelepiped is also able to be calculated through the orthogonalized candidate but with respect to $\boldsymbol{v}_{i_1}$ and $\boldsymbol{v}_{i_2}^\perp$. In general, this orthogonalization is the Gram-Schmidt process \cite{trefethen1997numerical}, formulated by 
\begin{equation}
\boldsymbol{v}_{i_k}^\perp=
\begin{cases}
\boldsymbol{v}_{i_k}, & k=1\\
\mathcal{T}_{I_{1:k}}(\boldsymbol{v}_{i_k}), & \text{otherwise}
\end{cases},
\end{equation}
where 
\begin{equation}
\mathcal{T}_{I_{1:k}}(\boldsymbol{v}_{i_k}) \coloneqq \boldsymbol{v}_{i_k} - \sum_{i \in I_{1:k}}\frac{\langle \boldsymbol{v}_{i_k}, \boldsymbol{v}_{i}^\perp \rangle}{\langle \boldsymbol{v}_{i}^\perp, \boldsymbol{v}_{i}^\perp \rangle}\boldsymbol{v}_{i}^\perp.
\end{equation}

In particular, when there is no sliding window, we can reuse the orthogonalization results of candidate items, which is equivalent to one-step modified Gram-Schmidt (MGS) orthogonalization \cite{trefethen1997numerical}. With this trick, our greedy inference algorithm only has $O(NTd)$ time complexity with extra $O(1)$ space. $t \le w$ is also equivalent to no sliding window case. The algorithm is summarized in Algorithm~\ref{alg:no_sliding_ssd}.

\begin{algorithm}
\SetAlgoLined
 Hyperparameter: return sequence length $T$ and trade-off coefficient $\gamma > 0$\;
 Input: qualities $\{r_i\}_{i=1}^N$ and embeddings $\{\boldsymbol{v}_i\}_{i=1}^N$\;
 Initialize: $t = 1, i_t = \arg\max_{j \in Z} r_j$\;
 $\mathcal{V} = \gamma \Vert \boldsymbol{v}_{i_t} \Vert$ \tcp*{Volume}
 \While{$t < T$}{
  \For{$j \in Z \setminus I_{1:t+1}$}{
  	$\boldsymbol{v}_j = \boldsymbol{v}_j - \frac{\langle \boldsymbol{v}_j, \boldsymbol{v}_{i_t} \rangle}{\langle \boldsymbol{v}_{i_t}, \boldsymbol{v}_{i_t} \rangle}\boldsymbol{v}_{i_t}$ \tcp*{One step MGS}
  }
  $t$++\;
  $i_t = \arg\max_{j \in Z \setminus I_{1:t}} r_j + \Vert \boldsymbol{v}_j \Vert \mathcal{V}$\;
  $\mathcal{V} = \Vert \boldsymbol{v}_{i_t} \Vert\mathcal{V} $
 }
 Output: item sequence $\{i_1, i_2, ..., i_T\}$
 \caption{SSD without sliding window}
 \label{alg:no_sliding_ssd}
\end{algorithm}

Let's move to time steps $t > w$, where we need to consider multiple windows. For simplification, let's first consider $t = w + 1$. A possible way that follows the strategy before is to orthogonalize $I_{2:w+1}$ starting from ${i_2}$. However, this way fully ignores item $i_1$ as well as the first window, which is against the SSD that measures the entire diversity by combining multiple windows. Therefore, we perform MGS in the current window while reserve the orthogonalized results of all selected items at the same time, allowing the current window inherit the information about previous windows. As a result, in time step $t$, the greedy inference consider the following objective:
\begin{equation}
\label{eq:ssd_obj_greedy}
\max_{j \in Z \setminus I_{1:t}} r_j + \gamma \Vert \mathcal{T}_{I_{l:t}}(\boldsymbol{v}_{j})\Vert\prod_{i \in I_{1:t}} \Vert \boldsymbol{v}_{i}^\perp\Vert,
\end{equation}
where $l=\max(1, t - w + 1)$. Note that the objective above maintain the orthogonalization in $\mathcal{X}$'s first order. Since $\mathcal{X}$ is symmetry on the first two orders, Eq.~\eqref{eq:ssd_obj_greedy} ensures that after the orthogonalization, an arbitrary set of consecutive $w$ elements in the same column of either the first and the second order is orthogonal.

Compared with no sliding window case, we need extra revert operations. For example, in time step $w + 1$, we need to revert the projections on $i_1$ for all remained candidates in $Z \setminus I_{1:w+1}$. These revert operations require $O(NTd)$ time complexity with extra $O(wN)$ space to store these projections. Hence, the total time complexity is still $O(NTd)$. To further speed up the real implementation, we further propose to apply circular queue trick to reduce memory copies \cite{cormen2009introduction}. The algorithm is summarized in Algorithm~\ref{alg:sliding_ssd}.

\begin{algorithm}
\SetAlgoLined
 Hyperparameter: return sequence length $T$, trade-off coefficient $\gamma > 0$ and sliding window size $w > 1$\;
 Input: qualities $\{r_i\}_{i=1}^N$ and embeddings $\{\boldsymbol{v}_i\}_{i=1}^N$\;
 Initialize: $t = 1, i_t = \arg\max_{j \in Z} r_j$\;
 $\mathcal{V} = \gamma \Vert \boldsymbol{v}_{i_t} \Vert$ \tcp*{Volume}
 Let $\mathcal{B}$ be a circular queue with size $w$ \tcp*{Bases}
 Let $\mathcal{P}$ be a circular queue with size $w$ \tcp*{Projections}
 \While{$t < T$}{
  \If{t > w}{
    $i = \mathcal{B}\text{.head}$\;
    $\mathcal{P}^{(i)} = \mathcal{P}\text{.head}$\;
    \For{$j \in Z \setminus I_{1:t+1}$}{
      $\boldsymbol{v}_j = \boldsymbol{v}_j + \mathcal{P}_j^{(i)} \boldsymbol{v}_{i}$ \tcp*{Revert}
    }
  }
  Push $i_t$ into $\mathcal{B}$\;
  Push an empty element with size $N$ into $\mathcal{P}$\;
  $\mathcal{P}^{(i_t)} = \mathcal{P}\text{.tail}$\;
  \For{$j \in Z \setminus I_{1:t+1}$}{
    $\mathcal{P}^{(i_t)}_j = \frac{\langle \boldsymbol{v}_j, \boldsymbol{v}_{i_t} \rangle}{\langle \boldsymbol{v}_{i_t}, \boldsymbol{v}_{i_t} \rangle}$ \tcp*{Projections on item $i_t$}
  	$\boldsymbol{v}_j = \boldsymbol{v}_j - \mathcal{P}^{(i_t)}_j \boldsymbol{v}_{i_t}$ \tcp*{One step MGS}
  }
  $t$++\;
  $i_t = \arg\max_{j \in Z \setminus I_{1:t}} r_j + \Vert \boldsymbol{v}_j \Vert \mathcal{V}$\;
  $\mathcal{V} = \Vert \boldsymbol{v}_{i_t} \Vert \mathcal{V}$
 }
 Output: item sequence $\{i_1, i_2, ..., i_T\}$
 \caption{SSD with sliding window}
 \label{alg:sliding_ssd}
\end{algorithm}

\subsection{Analysis}

We have introduced SSD for diversified feed recommendation and a corresponding greedy inference algorithm before. SSD regards the item sequence as a time series observed by the user, combining multiple sliding windows in the entire sequence to align user's perception when browsing the feed. This viewpoint is one of the key differences from other works on diversified recommendation. SSD uses the volume of a third-order tensor consists of multiple windows to define diversity. The most related idea is DPP \cite{chen2018fast}. In this section, we first prove that without the sliding window, the diversity term in SSD is the same as the one in DPP. We then analyze the differences between these two methods. Finally, we compare the complexity of greedy inference between SSD and DPP in either with or without sliding window case.

When the diversity is only required within a single window, consider the SVD decomposition for $\boldsymbol{X}_t = \boldsymbol{U} \boldsymbol{\Sigma} {\boldsymbol{V}}^\textsf{T} \in \mathbb{R}^{t \times d}$, where the $i$-th row of $\boldsymbol{X}_t$ is $\boldsymbol{v}_{i_t}$. Then we can obtain that
\begin{equation}
\det{\boldsymbol{X}_t{\boldsymbol{X}_t}^\textsf{T}} = \det{\boldsymbol{U} \boldsymbol{\Sigma} {\boldsymbol{V}}^\textsf{T}\boldsymbol{V} \boldsymbol{\Sigma} {\boldsymbol{U}}^\textsf{T}} = (\det{\boldsymbol{\Sigma}})^2.
\end{equation}
Note that $\boldsymbol{X}_t{\boldsymbol{X}_t}^\textsf{T}$ is the kernel matrix in DPP when ignoring qualities, and the right hand side is the squared diversity term in SSD's objective. The determinant of a matrix means the volume in geometry, i.e., DPP also uses the volume of the $t$-dimensional hyper-parallelepiped spanned by $I_{1:t+1}$ to represent the diversity. Therefore, without sliding windows, SSD and DPP have the same definition on the diversity. In sliding window case, SSD generalizes this volume to third-order tensors to combine multiple windows, which we believe more align with users' perception in feed recommendation.

The main differences between DPP and SSD are two-folds. 1) DPP is a powerful probabilistic technique from physics to measure the diversity, where the selection probability of a subset of items is proportional to the determinant with respect to the kernel matrix. There is no order concept in DPP's assumption. In other words, it is the greedy inference procedure that provides the sequence in DPP's recommendation. For long sequences, it is necessary to consider a sliding window to meet users' perception. However, DPP does not provide the way to combine multiple windows, and entirely ignoring out-of-window information is too rigorous. SSD instead regards the feed as a time series, involving both the order and the sliding window in the trajectory tensor. 2) In DPP's framework, the selected probability is proportional to the volume, where the quality plays as a multiplier on the embedding of the corresponding item. This combination is interpretable in geometry, where the volume increases monotonically with the quality. In other words, quality is a kind of enhancer over the natural embedding in DPP. For feed recommendation, recommenders exploit users' past interactions at the most time while exploring their interests a little. Inspired by this behavior, SSD instead considers that diversity plays an explorer under the quality with the exploit-explore strategy.

Recall that \cite{chen2018fast} proposed a fast greedy inference algorithm for DPP, where a kernel matrix $\boldsymbol{K}$ is required. For a large-scale recommender, it is impractical to pre-compute $\boldsymbol{K}$ with dense item embeddings in offline. During online serving, $\boldsymbol{K}$ requires $O(N^2d)$ time complexity and extra $O(N^2)$ space. Recall that, without sliding window, SSD returns the same result with DPP, but in our proposed greedy inference algorithm, $\boldsymbol{K}$ is unnecessary. Besides, in sliding window case, we do the orthogonalization only with the last selected item, making the time complexity independent of the window size $w$. The complexity comparisons are summarized in Table ~\ref{tab:complexity_table}. Under greedy inference, SSD is more efficient than DPP.

\begin{table}
\caption{Comparisons of greedy inference complexity for SSD and DPP \cite{chen2018fast} with dense item embeddings. In general, we have $N > T > w$ and $d > w$. }
\label{tab:complexity_table}
\begin{minipage}{\columnwidth}
\begin{center}
\begin{tabular}{ |c|c|c| }
\hline
Algorithm & Time Complexity & Space Complexity \\ \hline\hline
SSD without window & $O(NTd)$ & $O(Nd)$ \\ \hline
DPP without window & $O(NTd + N^2d)$ & $O(Nd + N^2)$ \\ \hline
SSD with window & $O(NTd)$ & $O(Nd)$ \\ \hline
DPP with window & $O(NTdw + N^2d)$ & $O(Ndw + N^2)$ \\ \hline
\end{tabular}
\end{center}
\end{minipage}
\end{table}

\section{Item Embedding}
\label{sec:item_embedding}
The SSD method introduced in the last section requires that items are embedded in an inner product space. Any item $i$ is represented by an embedding vector $\boldsymbol{v}_i$ and its similarity to any other item $j$ is computed as $\langle \boldsymbol{v}_i, \boldsymbol{v}_j \rangle$. Here, we describe how we compute the embedding vector for accurate item similarity measurement under long tail effect. We name our strategy as CB2CF although it has differences with~\cite{barkan2019cb2cf}.


\subsection{Model Architecture}
\begin{figure}
    \center
    \includegraphics[width=\columnwidth]{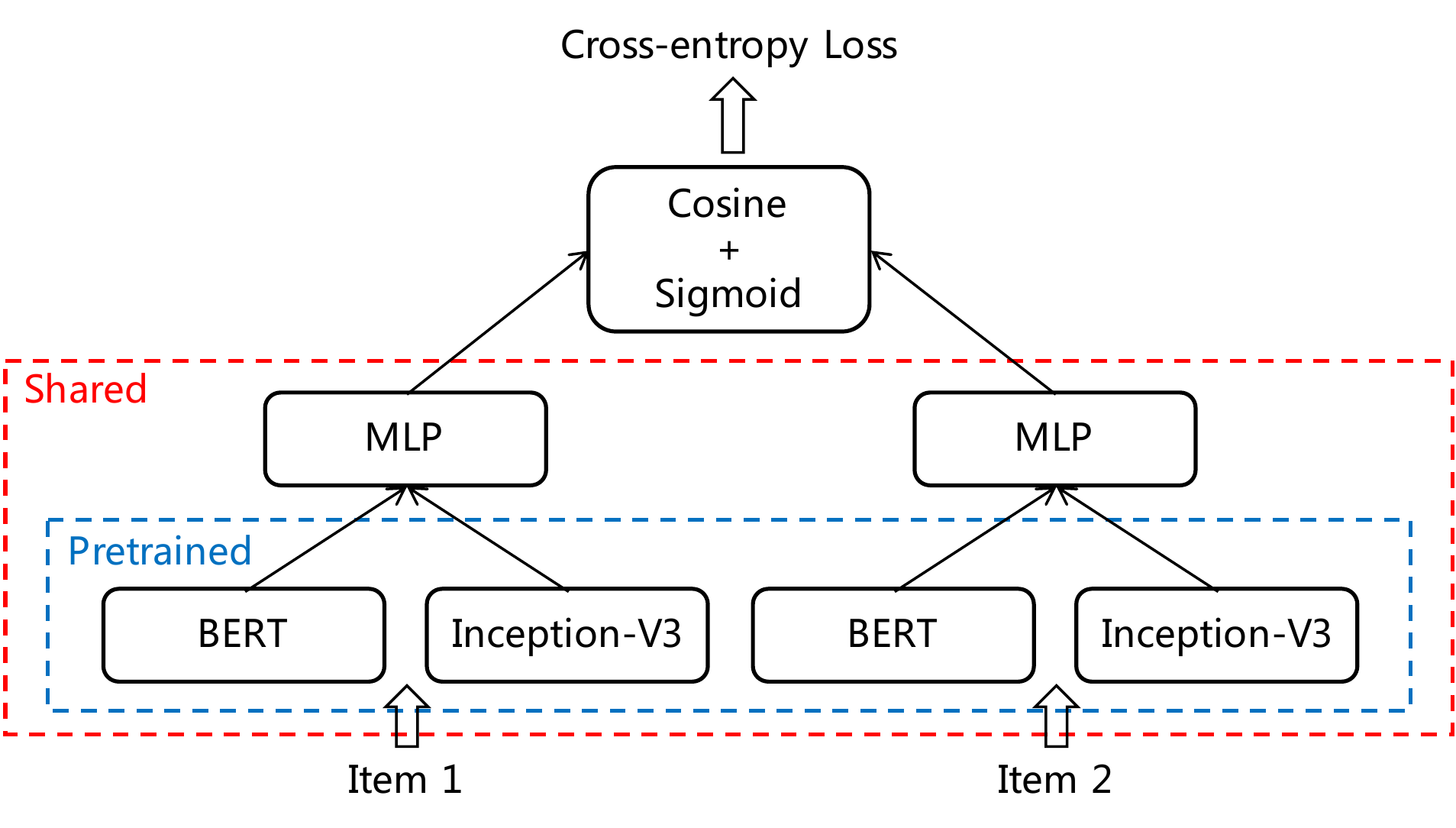}
    \Description[xiaohongshu-longtail]{A siamese structure for CB2CF.}
    \caption{The siamese network structure of the item embedding model.}
    \label{fig:network}
\end{figure}

The network structure of the embedding model is shown in Figure ~\ref{fig:network}. We adopt the siamese architecture~\cite{bromley1994signature}, which has been widely used in vision and natural language applications~\cite{neculoiu2016learning} for similarity computation. The siamese network is able to naturally learn the item embedding from pairs of items without the necessity to worry about deciding the order of items during training or selecting the branch of model at inference time. Given an item, we take the item description and cover image as input and created a text representation and image representation respectively for each of them with the help of pre-trained models. For text description, we adopt BERT~\cite{devlin2018bert} as the base model and follow the standard process to take the hidden state corresponding to the first token as the embedding. For the cover image, Inception-V3~\cite{szegedy2016rethinking} model is used to obtain the image representation. Then, we merge the representation of them by concatenating their corresponding representation, followed by a fully connected output layer into the target embedding $\boldsymbol{v}$. Finally we compute the cosine distance between the embeddings of the paired items. We will explain the adoption of cosine distance in Section~\ref{sec:item_embedding_normalization}. We treat this problem as a binary classification task and use the cross entropy loss as the objective.

\subsection{Training Data}
Next we discuss how we collect training samples for the embedding model. First, only item contents, such as textual description, cover image, and video's cover image are used as features in the embedding model. User engagements are not used as features. By doing so, the model is forced to infer similarity purely from item content. This prevents the model from being biased by the long tail effect, at least at the feature level, because items with no or very few user engagements can still be reliably computed by the model with sufficient content features to obtain their embedding vectors.

Second, for any item $i$ that has user engagements, if item $j$ is retrieved through ItemCF method \cite{linden2003amazon} with item $i$ as the seed item and item $j$ gets enough exposure in the final recommendation result, we have high confidence that item $i$ and $j$ is very similar from users' perspective. ItemCF ensures a basic degree of user-perceived similarity between $i$ and $j$ based on user engagements. If $j$ survives the recommender system many times to get exposure for many users, it is very likely that $j$ is similar to $i$ for many users. We treat these item pairs $<i,j>$ as positive samples. 

This sample set is not much affected by the long tail effect because even items with very few user engagements can serve as the seed item $i$. The item $j$, though, is more likely to be head items because our strategy requires it to have some minimal exposures. This is also why we do not require any user engagements on $j$ because it will further increase the probability of $j$ being head items. One upside of the design is that it enables the possibility to establish similarity relation between long tail and head items. This can help the embedding model to generalize from item content to infer similarity, even for long tail items.

Finally, for any item $i$ that has user engagements, we randomly sample an item $j^\prime$ retrieved from ItemCF method and use these item pairs $< i, j^\prime >$ as negative samples. If we allow $j^\prime$ to be any random item, we found in practice that the negative samples become too easy for the embedding model and it stops learning too early. All the above training samples can be automatically collected from existing recommender system easily. Our method is different from the original CB2CF work \cite{barkan2019cb2cf} in that we do not map item contents into CF vectors \cite{rendle2012bpr}. Instead, we force the model to learn from item content to infer a strengthened similarity signal produced by the ItemCF method.

\subsection{Normalization}
\label{sec:item_embedding_normalization}
As mentioned before, SSD generalizes the volume of the parallelogram spanned by the selected items' embeddings to sliding window case, matching users' perception in feed recommendation. When using inner product as the distance in CB2CF training, the L2-norm of an item is unbounded. This unboundedness may lead to serious unfairness among items because it offers uncontrollable contributions to the volume. Hence, we specify cosine similarity as the distance in CB2CF training, making all embeddings have equal L2-norm, i.e., lie on the high-dimensional unit ball's surface.

Another issue is the mismatch between the cosine distance and the volume. Volume is usually defined on the inner product space, where two vectors are entirely different if their inner product is 0. With cosine distance, the entirely different pair instead has value of -1. We address this issue by appending an extra dimension with element 1 after the normalization:
\begin{equation}
\label{eq:embedding_preprocess}
\boldsymbol{v} \coloneqq [\frac{\hat{\boldsymbol{v}}}{\Vert \hat{\boldsymbol{v}} \Vert}, 1],
\end{equation}
where $\hat{\boldsymbol{v}}$ is the output of CB2CF. This transformation keeps the similarity described by the cosine and achieves fairness and geometric property in SSD.



\section{Experiments}

In this section, we first introduce our implementation details in Section \ref{sec:implementation}, focusing on a real industrial recommender. Then, for item embeddings we show that CB2CF embedding is able to obtain accurate similarity measurement for long tailed items with offline evaluations in Section ~\ref{sec:offline_exp}. Finally, we provide the online A/B test comparisons between SSD and DPP in Section ~\ref{sec:online_exp}.


\subsection{Implementation Details}
\label{sec:implementation}
Here, we briefly discuss the implementation details in our experiments. In CB2CF embeddings, we mainly have two kind of content: text and image. For text, we pre-trained a BERT base model with around 40 million posts published on Xiaohongshu following the training scheme described in~\cite{liu2019roberta}. The sequence length is set to $128$. For image, we use the Inception-V3 model pre-trained on Xiaohongshu's internal image classification dataset. To ensure accurate positive labels, we collected two month's users' feedback on items that are retrieved through classical item-to-item CF technique~\cite{linden2003amazon}. We only consider an item that has been exposed to at least 100 users as positive. We randomly sample 400,000 pairs from this dataset as final positive set. In order to avoid the bias from most popular items, we make sure each post appears only once in this set. We randomly selected posts from the impressions to form a negative labelled set of the same size as the positive ones. The embedding dimension is set to $d=64$ for efficient inference (after preprocess in Eq. \eqref{eq:embedding_preprocess}, $d=65$). In addition, we pre-compute the embeddings of all available items offline for online serving.

Through some offline simulation and online analysis, we find two tricks that can further improve the performance. 1) In practice, the quality is usually trained to regress a combination value of multiple business goals, such as engagement and watch time. Users often have different ranges of values, leading to inconsistent trade-offs for a fixed $\gamma$. An ideal solution is to make $\gamma$ depend on the user. We find that standard normalized quality scores on all candidate items in a single request work quite well. 2) We observed that in greedy inference procedure during the construction of the sequence, the diversity values drop more quickly than the quality values when the user has concentrated interests. These significantly different trends introduce a stability issue on the trade-off coefficient. We find that the following approximate objective improves the stability a lot in practice:
\begin{equation}
\label{eq:ssd_obj_greedy_fix}
\max_{j \in Z \setminus I_{1:t}} r_j + \gamma\Vert \mathcal{T}_{I_{l:t}}(\boldsymbol{v}_{j})\Vert.
\end{equation}
Note that although different from Eq.~\eqref{eq:ssd_obj_greedy} in SSD, Eq.~\eqref{eq:ssd_obj_greedy_fix} inherits previous windows' information in orthogonalized items $I_{l:t}$. We call this method $\text{SSD}^*$ which makes SSD more stable with regard to the coefficient $\gamma$ and improves SSD's effect even further in our online test in Section \ref{sec:online_exp}.

Beyond quality and diversity, a real-world recommender needs to simultaneously satisfying business constraints, e.g., limit the number of items associated with plastic surgery. These constraints are often ensured in the tail of the recommendation pipeline, e.g., \cite{wilhelm2018practical} implement them after a DPP optimizer in YouTube. We argue that in most cases, we can know whether an item satisfy a constraint in $O(1)$ time, and the number of constraints is usually less than $d$. Therefore, with the same time complexity, we can reduce the conflicts between business constraints and diversified recommendation by removing items that do not satisfy business constraints from $Z \setminus I_{1:t}$ in time step $t$. We applied this strategy in all our online experiments.

\begin{table}
  \caption{Analysis of most similar items computed using CF and CB2CF for long tailed items.}
  \label{tab:cf_taxonomy_analysis}
  \resizebox{\columnwidth}{!}{
  \begin{tabular}{c|c|c|c|c|c}
    \toprule
    model &\multirow{2}{*}{\shortstack{Taxonomy \\ of Seeds}} & \multicolumn{4}{c}{Taxonomy of top 10 most similar items} \\\cline{3-6}
    & & \multicolumn{1}{c|}{Sport} 
    & \multicolumn{1}{c|}{Fashion} 
    & \multicolumn{1}{c|}{Travel} 
    & \multicolumn{1}{c}{Food}  \\
    \hline
    \multirow{4}{*}{CF} & Sport & \textbf{43.9\%} & 7.9\% & 3.4\% & 5.3\% \\
    & Fashion & 1.1\% & \textbf{47.0\%} & 5.5\% & 4.3\%\\
    & Travel & 2.9\% & 10.6\% & \textbf{30.8\%} & 13.1\%\\
    & Food & 0.9\% & 6.5\% & 6.4\% & \textbf{48.0\%} \\
    \hline
    \multirow{4}{*}{CB2CF} & Sport & \textbf{74.5}\% & 1.5\% & 1.9\% & 1.2\% \\
     & Fashion & 0.2\% & \textbf{86.9\%} & 1.4\% & 0.5\% \\
     & Travel & 0.9\% & 4.2\% & \textbf{71.6\%} & 5.7\% \\
     & Food & 0.1\% & 0.6\% & 2.3\% & \textbf{87.7\%} \\
  \bottomrule
\end{tabular}
}
\end{table}

\begin{figure}
    \center
    \includegraphics[width=\columnwidth]{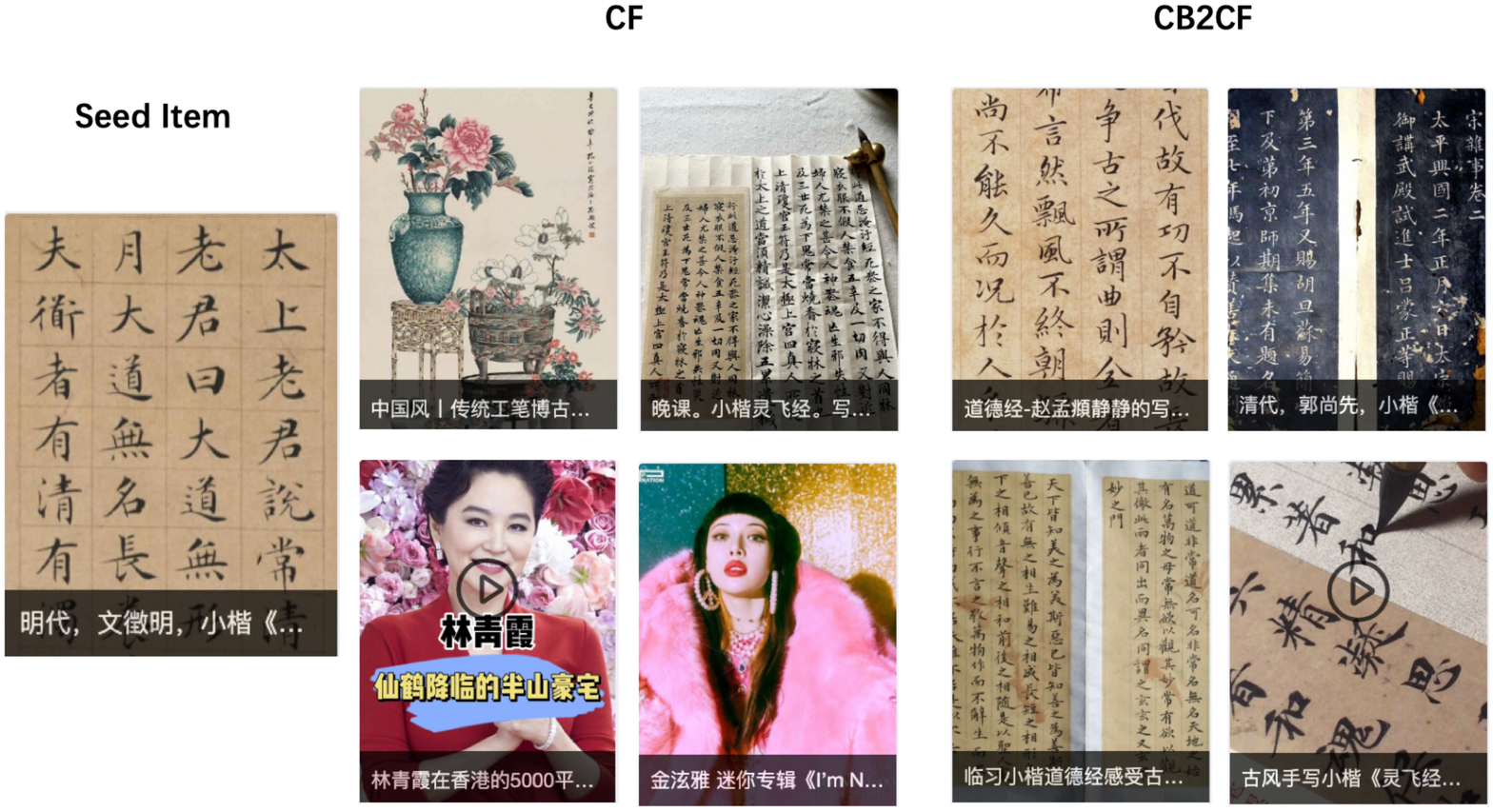}
    \caption{Most similar items produced by CF and CB2CF.}
    \Description{Most similar items produced by CF and CB2CF in Xiaohongshu dataset.}
    \label{fig:cf_cb2cf_demo}
\end{figure}

\subsection{Offline Evaluation}
\label{sec:offline_exp}
Here, we conduct an offline experiment to understand better the difference of item similarity computation using CF technique and the proposed CB2CF item embedding for long tailed items. 

We restrict our attention to long tailed items with only few numbers of engagements. We randomly sampled 100,000 seed items, which all have less than 3 engagements in a month and obtain the top 10 most similar items scored by ItemCF \cite{linden2003amazon}. For CB2CF method, the item embedding for the same set of items are computed and the most 10 similar items for each seed item are retrieved within the item embedding space. In order to compare the similarity of these retrieved items, we label each item with a taxonomy tag, such as sport, fashion, food, etc., with our internal taxonomy system to denote the category of the item. Here we only use the taxonomy as a coarse-grained approximation to compare the similarity of the items. In Table~\ref{tab:cf_taxonomy_analysis} we report the distribution for some popular taxonomies of the seed items and the retrieved ones. The items from these four taxonomies are generally quite dissimilar from each other. It is evident that for long tailed items our proposed CB2CF method is more effective in terms of measuring similarity for long tail items. In addition, Figure~\ref{fig:cf_cb2cf_demo} depicted an example of the most similar items generated from CF and CB2CF given a low engaged item which describes Chinese calligraphy. Clearly, the proposed CB2CF method manages to retrieve more similar items compared to CF which retrieves entertainment and painting related items under this scenario.

\begin{table}
  \caption{Improvement in online A/B tests over DPP \cite{chen2018fast}.}
  \label{tab:ab_table}
  \begin{minipage}{\columnwidth}
  \begin{center}
  \begin{tabular}{c|c|c|c|c}
    \toprule
     & Time spent & Engagements & ILAD & MRT \\
    \midrule
    SSD & +0.25\%  & +0.71\% & +0.37\% & +0.53\% \\
    $\text{SSD}^*$ & +0.42\%  & +0.81\% & +0.32\% & +0.68\% \\
  \bottomrule
  \end{tabular}
  \end{center}
  \end{minipage}
\end{table}

\subsection{Online A/B Test Result}
\label{sec:online_exp}
In this subsection, we introduce our online A/B test result for SSD and $\text{SSD}^*$. The Explore Feed's recommender system in Xiaohongshu follows the structure shown in Figure \ref{fig:structure_recsys}. In the policy module, the number of items considered is $N=600$, and the returned number of items is $T=80$. The proper value of window size $w$ usually depends on the application. In our setting, $w$ will be no more than several times of the maximal number of items that can fit in a screen and is chosen from  $\{8, 10, 12, 15, 20\}$. In this experiment, we used grid search to find the optimal window size $w$ and the coefficient $\gamma$ for DPP, SSD and $\text{SSD}^*$ respectively and only best performance of the methods are compared and reported here.

For the control group, we randomly selected 10\% users of Xiaohongshu and applied the DPP greedy inference method \cite{chen2018fast}. For each treatment group, a randomly selected 10\% users were applied the SSD or $\text{SSD}^*$ method with greedy inference algorithm in Section \ref{sec:greedy_inference}. DPP, SSD and $\text{SSD}^*$ all use the same item embedding introduced in Section \ref{sec:item_embedding} in all online experiments.

We use four metrics to measure the online performance: time spent on feed, the number of engagements, the intra-list average distance (ILAD)\cite{zhang2008avoiding} and the mean read taxonomies (MRT). The latter two are defined as
\begin{displaymath}
\begin{aligned}
\text{ILAD} &\coloneqq \underset{u \in U}{\text{mean}} \underset{i,j \in \text{Imp}_u, i \ne j}{\text{mean}} (1 - S(i, j))\\ 
\text{MRT} &\coloneqq \underset{u \in U}{\text{mean}} |\{T(i)|i \in \text{Clk}_u\}|
\end{aligned}
\end{displaymath}
where $U$ is the set of all users, $\text{Imp}_u$ is the set of all items recommended to user $u$ and $\text{Clk}_u$ is the set of all items that user $u$ clicked. $S(i, j)$ represents the similarity between item $i$ and item $j$, and $T(\cdot)$ maps an item into its taxonomy. ILAD 
measures diversity using the average pair-wise dissimilarity among user viewed items. MRT describes the diversity using the average number of item categories that users clicked. Both are commonly used diversity metrics in industry recommender systems. 

The online A/B test result averaging over 7 days experiment period is shown in Table~\ref{tab:ab_table}. SSD significantly improves the diversity and quality of the recommendation result  over DPP. Besides, SSD reduces 7ms on the 99th percentile (P99) of latency and reduces relatively 27\% memory requirement of the policy module. For $\text{SSD}^*$, we can see that it further improves SSD's overall performance.

\section{Conclusion}
Currently, many social media platforms recommend a sequence of items for users to browse and engage with, in a type of product called feed. In this paper, we propose to study recommendation diversity problem from an item sequence perspective using time series analysis techniques in this scenario. We derive a method called sliding spectrum decomposition (SSD) in this direction that better captures users' perception of diversity in long sequence scenario. We also share our experiences on using CB2CF strategy to obtain an item embedding that can accurately measure item similarity under long tail effect. The effectiveness and efficiency of the proposed methods are demonstrated and verified through theoretical analysis, offline experiments and online A/B tests.

\bibliographystyle{ACM-Reference-Format}
\bibliography{bibliography}

\appendix

\end{document}